\documentclass[11pt,twoside]{article}
\usepackage{asp2010}
\usepackage{epstopdf}
\usepackage{graphicx}

\resetcounters

\begin{document}

\title{Implications of Radial Migration for Stellar Population Studies}
\author{
Rok Ro\v{s}kar$^{1,2}$, 
Victor P. Debattista$^3$, 
Sarah R. Loebman$^1$,
\v{Z}eljko Ivezi\'{c}$^1$,
Thomas R. Quinn$^1$}
\affil{$^1$Astronomy Department, University of Washington, Box 351580 Seattle, WA, 98195, USA}
\affil{$^2$Institute for Theoretical Physics, University of Z\"{u}rich, CH-8057 Z\"{u}rich, Switzerland}
\affil{$^3$Jeremiah Horrocks Institute, University of Central Lancashire, Preston, PR1 2HE, UK}

\begin{abstract}
Recent theoretical work suggests that it may be common 
for stars in the disks of spiral galaxies to migrate radially 
across significant distances in the disk. Such migrations 
are a result of resonant scattering with spiral arms and 
move the guiding centers of the stars while preserving the 
circularities of their orbits. Migration can therefore efficiently 
mix stars in all parts of the Galactic disk. We are rapidly 
approaching an important confluence of theory and 
observation, where we may soon be able to uncover 
signatures of such processes in our own Milky Way. The 
resolution and robustness of the physical modeling in 
simulations has improved drastically, while observational 
datasets are increasing in depth and astrometric accuracy. 
Here, we discuss the results from our idealized N-body/SPH 
simulations of disk formation and evolution, emphasizing 
specifically the observational consequences of stellar 
migration on the solar neighborhood and the vertical 
structure of the disk. We demonstrate that radial mixing of 
stars is a crucial dynamical process that we must try to 
understand if we are to draw significant conclusions about 
our Galactic history from the properties of stars in our vicinity.
\end{abstract}

\section{Introduction}
Disks of spiral galaxies are kinematically cool,
rotationally-supported, self-gravitating systems made up of stars and
gas. Stars form in the thin gas layer and begin their lives
on mostly circular orbits due to the efficiency with which gas is able to
shed excess energy. The stellar orbits are subsequently 
heated through interactions with giant molecular clouds (GMCs),
external perturbations from infalling substructure, and resonances with disk structure. 
Observations of stars in the solar neighborhood show that velocity dispersion is
related to stellar age by a power-law \citep{Holmberg:2009}. Given
that the epicycle amplitude in the solar neighborhood is limited by
$\Delta R \simeq \sqrt{2} \sigma_R / \kappa$ (equation 3.99
\citealt{Binney:2008}, hereafter BT08), $\sigma_R \sim 50$~km/s
\citep{Holmberg:2009} for the oldest stars in the solar neighborhood,
and $\kappa_0 = 37$~km/s/kpc (BT08), the largest orbital excursions
for stars at the solar radius are $\lesssim 2$~kpc. Averaged over the
entire sample, the amplitude of radial oscillations $\Delta R \simeq
1.3$~kpc, allowing stars to traverse across $\sim2$ kpc during a
single orbit.

Such radial orbital oscillations of stars naturally affect the
distribution of stellar population properties in the solar
neighborhood.  If regions of the galaxy evolve approximately as
closed-box systems, then one expects there to be a tight relationship
between the age of a star and its metallicity, the so-called
age-metallicity relation (AMR) \citep{Twarog:1980}. Radial
oscillations of stars dilute such a relation, and have been considered
in the past to explain the large scatter in the observed AMR
\citep{Edvardsson:1993, Holmberg:2009}. However, if we naively assume
a relatively steady metallicity gradient over the past few Gyr of
$\sim 0.06$ dex kpc$^{-1}$ \citep{Daflon:2004}, the dispersion of
the stellar metallicity in the solar neighborhood arising solely from
such orbital excursions should be $\lesssim 0.1$~dex. The observed
dispersion in the AMR in the solar neighborhood is larger by at least
a factor of 2 \citep{Holmberg:2009}. Of course, the gradient could
have been steeper in the past, but correcting for orbital
eccentricities in the observed Geneva-Copenhagen sample does not
explain away the large dispersion in the AMR \citep{Nordstrom:2004,
Binney:2007}.

The observationally-limited maximum epicycle amplitude of $\lesssim
2$~kpc leads to a picture of galactic disks where stars must remain
relatively close to their radii of origin, only increasing their
epicyclic energies in response to perturbations in the disk. The need
for more substantial radial mixing was first discussed by
\citet{Wielen:1996}, who argued that the Sun could not have originated
at its present radius based on its anomalously high metallicity with
respect to the surrounding stars and the local ISM
metallicity. \citet[][hereafter SB02]{Sellwood:2002} spearheaded the
recent resurgence in the interest in radial mixing by showing that
radial migrations of even larger magnitude than that postulated by
\citet{Wielen:1996} are not only possible but very likely in the
presence of recurring transient spirals. SB02 demonstrated that the
migration takes place due to scattering at the corotation resonance of
the spiral, but the large changes in the stellar guiding centers are
not accompanied by significant heating. The lack of heating is important
because mostly circular stellar orbits are typically assumed to have had a relatively
quiet history - SB02 showed that this assumption is not necessarily
true. Subsequently, \citet{Lepine:2003} explored the orbital evolution of stars under the influence
of corotation resonance; \citet{Roskar:2008, Roskar:2008a} studied the effects of radial 
migration in a full $N$-body simulation of disk formation; \citet{Schonrich:2009, Schonrich:2009a} 
were the first to incorporate these ideas into a chemical evolution model of 
the Milky Way; meanwhile \citet{Debattista:2006} and \citet{Minchev:2010} investigated 
mixing and heating via chaotic bar-spiral coupling. 

We examine this paradigm further by means of self-consistent, 
though idealized in its initial configuration, simulation of disk 
galaxy formation. The results of this simulation have previously been 
reported in \citet[R08 collectively hereafter]{Roskar:2008, Roskar:2008a}
and \citet{Loebman:2010}.  
In this contribution, we specifically focus on the importance of radial 
migration for studies using local (solar neighborhood) stellar samples
to infer something about our Galaxy's past evolution (such as 
 \citealt{West:2008, Bochanski:2010}; see also Bochanski, this volume).

\section{The Simulation}
The details of our simulation method have been discussed previously 
in R08. Here, we recall the salient qualitative aspects. The basic picture
is one of a hot, spherical gaseous halo embedded in a massive dark matter 
halo, such as one might expect to exist after the last major merger for 
a Milky Way (MW) type galaxy. Thus, our model is
initialized with a spherical gas halo in hydrostatic equilibrium set 
by the potential well of a $10^{12}$M$_{\odot}$ dark matter (DM) halo. Both
halos follow the same NFW profile \citep{Navarro:1997}, but we impart
a spin upon the gas corresponding to a cosmologically-motivated dimensionless spin parameter 
$\lambda = 0.039$ \citep{Bullock:2001}. The two components are represented by 1 million
particles each - this results in a mass resolution of $\sim 10^5$M$_{\odot}$
for the gas and $\sim 10^6$M$_{\odot}$ for the DM. We evolve the system
with the parallel $N$-body + Smooth Particle Hydrodynamics code
GASOLINE \citep{Wadsley:2004} for 10 Gyr. As the simulation proceeds, 
the gas is able to cool and collapse to the center of the potential well, 
forming a centrifugally-supported disk. 

The crucial point here is that we do not insert by hand any properties 
of the disk. Instead, its final properties are a product of a complex interplay 
of various processes (gas cooling, stellar feedback, star formation, 
self-gravity).
Our simulation code allows the gas to form 
stars once appropriate temperature and density are attained, thus forming
a disk of stars self-consistently within our sub-grid star-formation framework. 
The stars in turn are modeled as evolving stellar populations, a fraction of 
which explode as supernovae (of type Ia and II), returning metals 
back to the ISM. We follow the yields of $\alpha$ elements and iron separately,
allowing us to track gross abundance patterns. 

Our model should therefore be regarded as a crude chemical evolution 
model coupled with self-consistent $N$-body dynamics. Since we do not
set any properties of the disk a priori, our simulation is not a fit to the Milky 
Way. However, the simulation yields a disk that agrees qualitatively in many 
ways with the properties of our Galaxy. A critical concern is that the amount of
structure that forms in the disk is unreasonable - should the amplitude and frequency
of spirals be unusually high, this could result in unreasonable heating rates and 
unrealistic rates of migration. Importantly, the heating rates 
as derived from the age-velocity dispersion relationship in the model 
are even somewhat lower than the MW - power law fits give indices of 
0.24, 0.21, 0.25, 0.37 for total, u, v, w dispersions respectively, compared with 0.40, 0.39, 0.40, and 0.53
for the same quantities derived for the solar neighborhood from the 
Geneva-Copenhagen survey \citep{Holmberg:2009}. We speculate that 
the discrepancy arises from the presence of additional heating sources 
in the Galaxy, which are not captured by the simulation such as 
substructure (our disk evolves in isolation) and giant molecular clouds. In our model, the heating is primarily
from spirals, and these numbers indicate that the amount of 
asymmetric structure is not unreasonable. Our simulated disk also yields
a reasonably flat rotation curve, though it flattens at $\sim 240$ km/s compared 
to 220 km/s for the MW indicating that the model is slightly more massive. 
The scale length of the simulated disk is 3.2 kpc compared to 2.6 kpc derived
for the MW from SDSS data \citep{Juric:2008}, and the gas fraction is $\sim10$~\%.

\section{Radial Migration in Disks}
On a Galactic scale, the solar neighborhood occupies a very small 
volume - the largest samples contain stars from a sphere only 
a few tens of parsecs in radius. It is therefore tempting to consider the 
grouping of stars around the Sun to be of a common origin, if not
of direct relation through a common birth cluster (i.e. the ``siblings'' of
the Sun \citealt{Portegies-Zwart:2009}), then at least by virtue of 
having been born in the same part of the Galaxy. The latter assumption 
is made by virtually every chemical evolution model of the MW
\citep[e.g.][]{Matteucci:1989, Carigi:1996, Boissier:1999,
Chiappini:2001}.

Such assumptions, as we show below, oversimplify the dynamical evolution of stars
in a galactic disk. Stellar orbits are profoundly affected by the 
spontaneous growth of spirals in a self-gravitating disk. The heating
offered by such transient spirals has been extensively studied in the 
literature, and is required to explain the observed heating 
rates (BT08). However, transient spirals can also
cause large changes in stellar orbital radii, while keeping the 
random energy of the orbit untouched, i.e. without heating 
\citep{Sellwood:2002}. This results in radial migration of 
stars, where individual radii in a MW-type disk can change
by several kpc during the lifetime of the disk.  

\begin{figure}[!ht]
\centering
\plotone[width=3in]{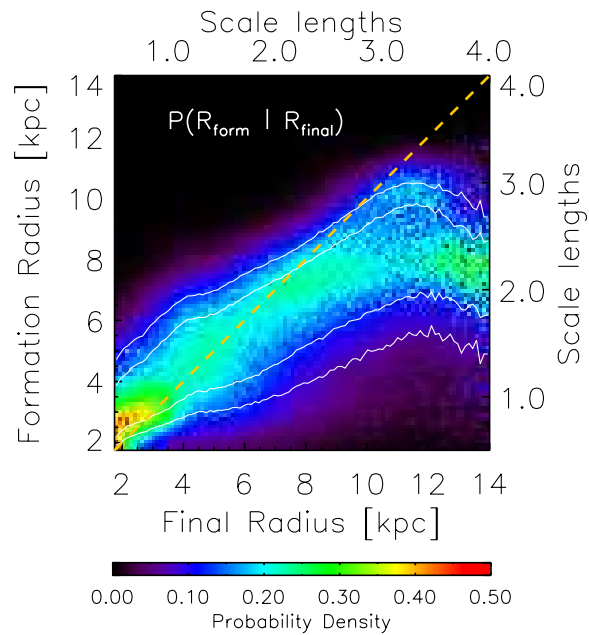}
\caption{Probability density plot of formation radius given a final radius for
stars in the simulation. The inner (outer) white contours enclose 50\% (75\%)
of the mass. If radial migration was unimportant, the highest probabilities would
concentrate along the dashed yellow 1:1 line.}
\label{fig:rform}
\end{figure}

Figure~\ref{fig:rform} shows the probability density plot of formation radii as a function
of final radii for the simulated disk at the end of the simulation. It is evident that the 
above assumption of stars found locally now being somehow related, is incorrect - 
whichever final radius one chooses, its stellar population is significantly contaminated
by stars from other radii. The extent of this contamination is drastic - at 8 kpc, 
a $\sim2\%$ chance exists that a star has formed at a radius as small as 2 kpc. 

\begin{figure}[!t]
\begin{minipage}{1.5in}
\includegraphics[width=1.8in]{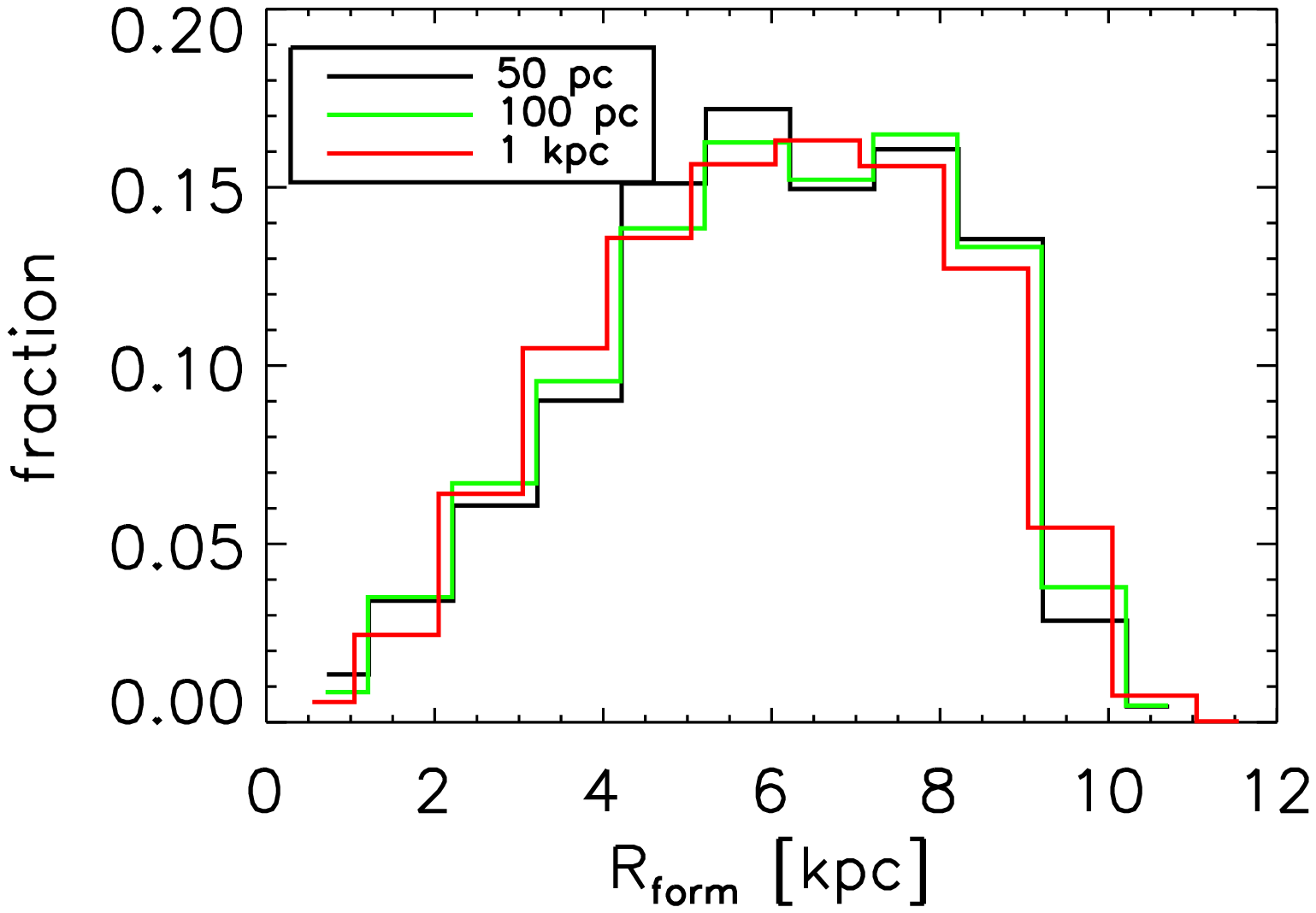}
\end{minipage}
\begin{minipage}{1.5in}
\includegraphics[width=1.8in]{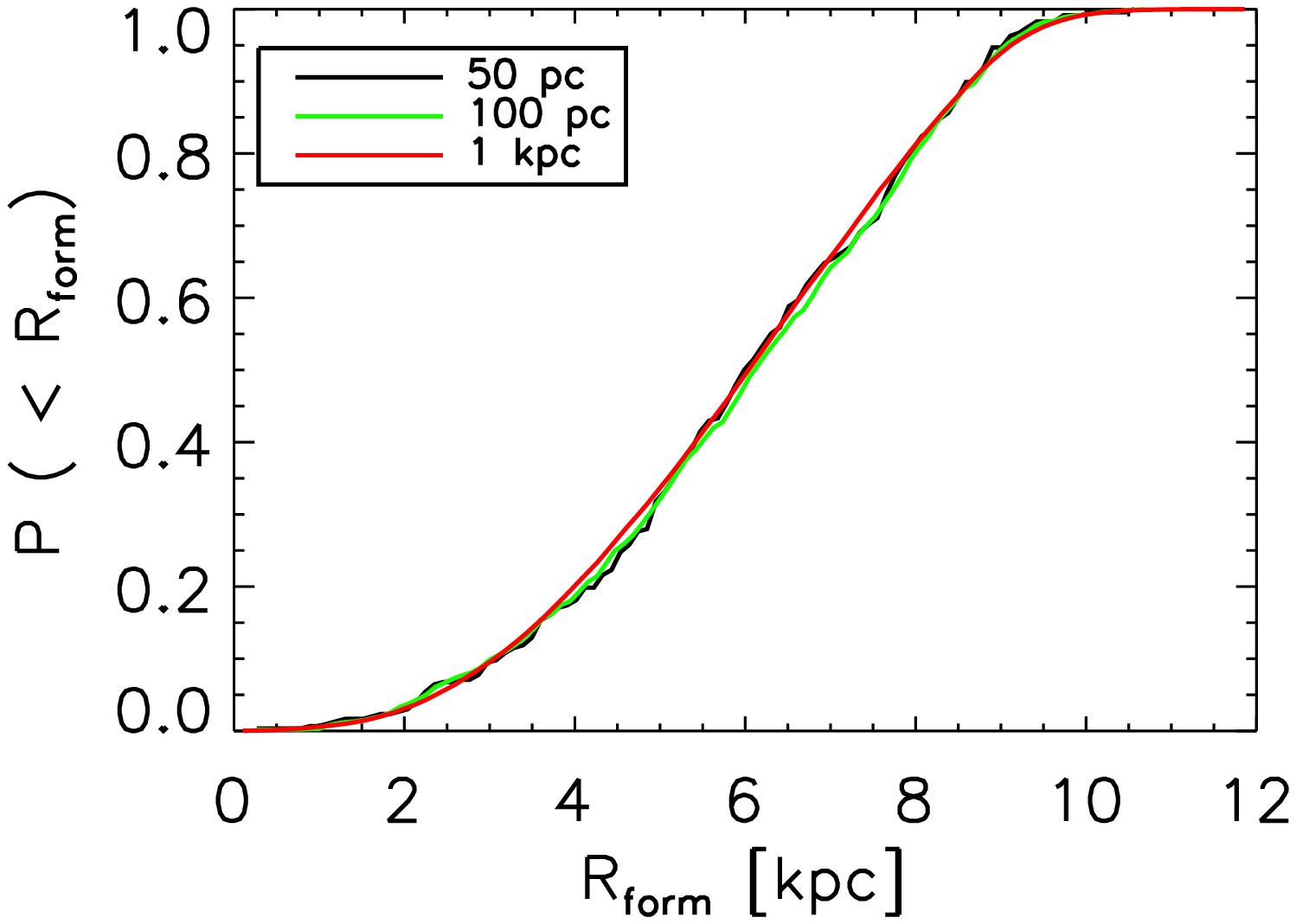}
\end{minipage}
\begin{minipage}{1.5in}
\includegraphics[width=1.8in]{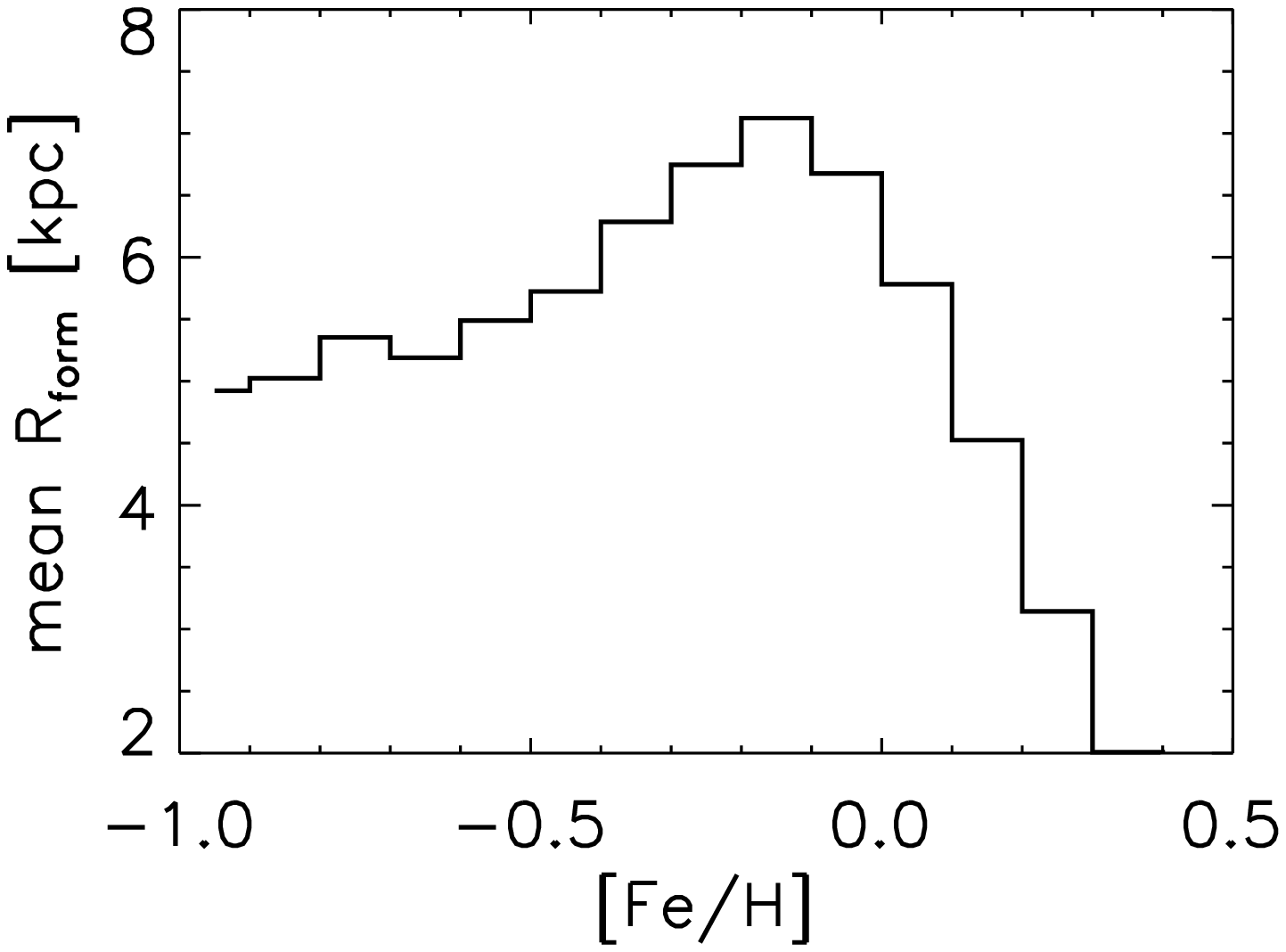}
\end{minipage}
\caption{{\bf Left:} Distribution of formation radii $R_{form}$ for stars found in three 
different annuli centered on 8 kpc. The black, green and red lines show the distribution for
annuli 50, 100, and 1000 pc wide respectively.
{\bf Center:} The cumulative distribution function of $R_{form}$ for the same annuli. More than
50\% of the stars come from $R < 6$~kpc. 
{\bf Right:} Mean $R_{form}$ as a function of metallicity. The ISM metallicity at 8 kpc is approximately Solar. 
}
\label{fig:rform_dist}
\end{figure}

For stellar population studies focusing on the solar neighborhood, the most 
important concern is how the narrow region around the Sun 
may be affected by such mixing. Due to the limited resolution of our simulation, 
we cannot isolate a single spherical region 50 pc in radius, but instead focus on the average
properties of a ``solar neighborhood'' defined as radial annuli centered on 8 kpc with 
several different $\Delta R$. Figure~\ref{fig:rform_dist} shows the distributions of formation
radii for these annuli. From the left panel, it is clear that the distribution is skewed to $R < 8$~kpc
and that it is quite extended. A better measure of the contribution of stars born at smaller radii to the overall
mix at the solar radius is given by the cumulative distribution function shown in the center panel. 
 At least 50\% of stars presently in the solar neighborhood have 
come from $R < 6$~kpc and $> 80\%$ of the particles currently in the solar neighborhood have come
from radii smaller than 8 kpc. 

In the right panel of Figure~\ref{fig:rform_dist}, we show the mean formation radii as a function 
of metallicity. Due to the fact that the disk grows from the inside-out, the majority of metal-rich stars therefore
come from the interior of the disk. As a result, the high-metallicity end of the stellar distribution preferentially
originates in the inner disk. Of course, because the local ISM metallicity is approximately solar, \emph{all} stars 
with metallicities above solar must have come from another part of the galaxy, unless significant recent 
infall of pristine gas diluted the metals or the ISM is azimuthally very inhomogeneous. 

Such distributions are of particular interest in light of the well-known 
planet-metallicity correlation for main-sequence FGK dwarfs
hosting gas giants \citep{Fischer:2005}. If the correlation is intrinsic (i.e. availability of planet-building material
in the protoplanetary nebula) rather than a result of the pollution of 
a star's atmosphere by the accretion of one of the planets, the right panel of Figure~\ref{fig:rform_dist} 
suggests that locally-found planet hosts should have spent considerable fractions of their lives in the 
interior of the Galaxy. Should the planet-metallicity correlation extend to hosts of lower-mass planets, 
the galactic history of such stars could have interesting consequences for 
the searches for signatures of extraterrestrial life (note that this correlation does not seem to exist
for planets around giants, e.g. \citealt{Pasquini:2007, Ghezzi:2010}).

\section{Vertical Evolution - Migrating into the Thick Disk}

\begin{figure}[!th]
\centering
\includegraphics[width=4in]{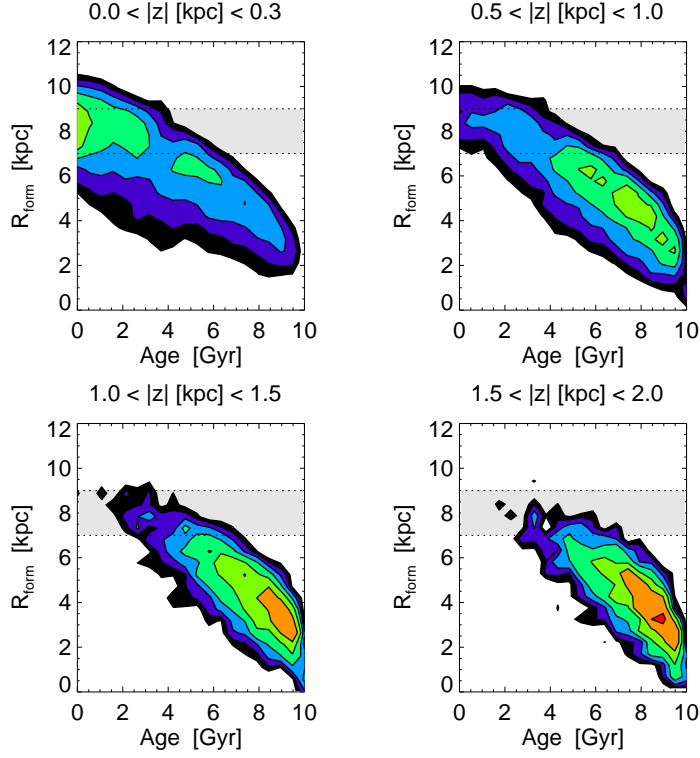}
\caption{Distributions of formation radius vs. age for particles found in various slices above the midplane,
in a 2 kpc annulus centered on 8 kpc (shown by the shaded region). Due to the high density 
of individual points, logarithmically-spaced contours are used as a proxy - colors indicate relative 
density, increasing from blue to red.}
\label{fig:rform_age}
\end{figure}

As stars move outward from the inner disk, they preserve their vertical energy, 
but the restoring force from the disk decreases due to the smaller surface density. 
Consequently, the amplitudes of their vertical oscillations also increase. The basic nature of this process 
is illustrated by Figure~\ref{fig:rform_age}. At the midplane (top left), the stellar population mix is 
strongly influenced by local star formation so the majority of stars are locally born and 
relatively young. Note, however, that the distribution of $R_{form}$ is very extended,
as shown in Figure~\ref{fig:rform_dist} (i.e. the projection of top left panel of Figure~\ref{fig:rform_age}
along onto the y-axis). As we consider vertical slices at increasing heights from the midplane, 
the stellar populations become dominated by old stars that have come from the inner disk. 
Fitting the vertical disk profile at 8~kpc with a double sech$^2$ profile yields two distinct components
with scale heights of 381 pc and 913 pc, roughly consistent with the values found for the thin 
and thick disk of the MW (270 pc and 1200 pc \citealt{Juric:2008}).

Because the disk grows and evolves, the birth environment
of these migrated stars was quite different from their present surroundings. The metallicity
gradient in the past was steeper (see Figure 2 of \citealt{Roskar:2008a}), resulting in a large number
of stars forming at low metallicity. The disk was also young and not yet polluted by supernova Ia metals
so these old stars tend to be $\alpha$-enhanced. As these stars are old, they have been 
heated through secular processes in the disk and therefore lag the local standard of rest. In short,
the migrated population of old stars at the solar radius has all of the characteristics of the thick disk.
See \citet{Schonrich:2009a} and \citet{Loebman:2010} for more detailed comparisons between the
properties of migrated stars and the observed MW thick disk. 
 
\begin{figure}[!ht]
\centering
\includegraphics[width=4.5in]{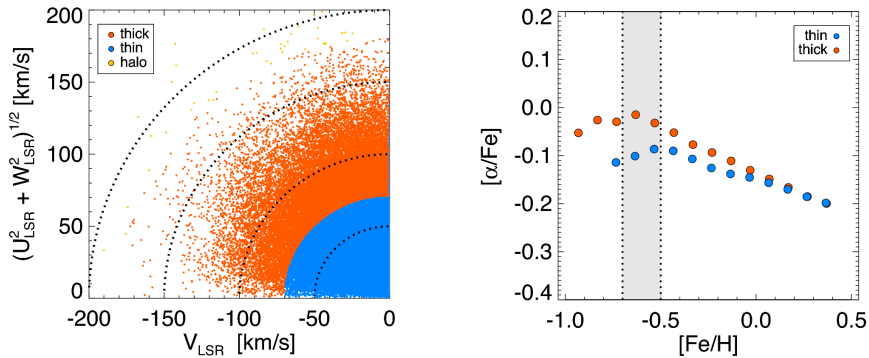}
\caption{{\bf Left:} Toomre diagram for stars in the midplane between 7-9 kpc. Color indicates a 
kinematic cut to define the thick and thin disks - stars with $V_{tot} \lesssim -70$~km/s are assigned thin disk 
membership, while stars with $-150 \lesssim V_{tot} < -70$~km/s are considered thick disk.
{\bf Right:} $\alpha$-element enrichment as a function of metallicity of stars selected to be in the thick and thin 
disks based on kinematics.}
\label{fig:thickdisk}
\end{figure}

In Figure~\ref{fig:thickdisk} we show a part of this comparison. The left panel shows the Toomre diagram
for the particles in the midplane at the solar radius. We select the thin and thick disk stars based on their
kinematics, as is frequently done in the literature (e.g. \citealt{Bensby:2003, Bensby:2005}), though because
we do not know the kinematic properties of different components a priori we use a simple ad-hoc hard-cut 
dividing the thin and thick disk at -70 km/s. Nonetheless, dividing the stars in this way gives us a rough idea 
of the interdependence between kinematics and chemistry. In the right panel of Figure~\ref{fig:thickdisk},
we show the $\alpha$-element enrichment as a function of metallicity for the kinematically-identified
thin and thick disk stars - the thick disk is enhanced, on average, in $\alpha$ elements, consistent
with observational results for the MW thick disk \citep{Bensby:2005}. 

Given the prevalence of sub-structure and streams found in the MW halo, presumably a consequence
of the hierarchical merging process integral to the LCDM paradigm of galaxy formation, it is 
usually assumed that the thick disk is likewise a relic our Galaxy's cosmological history (see 
\citealt{Wyse:2008} for a review). While we do not try to dispute the fact that our Galaxy is 
embedded in a tumultuous environment where disk-perturbing interactions are frequent, 
the fact remains that radial migration \emph{will} contribute to the stellar populations found away from the disk 
plane. Radial migration must therefore be considered when assessing the cosmological significance of the
stars found in the thick disk. 

\section{Conclusions}

In galaxies with recurring spirals, stars do not remain near their birth radii. The guiding centers of their
orbits can evolve drastically over the course of their lives. Conversely, no part of the disk in a spiral 
galaxy is free from contamination by stars that have come from other disk regions. Interpretations
of stellar population properties found anywhere in the MW disk need to account for this fact. We have
shown that in our models of MW-like disk formation, more than 50\% of the stars have come from 
$R \lesssim 6$~kpc. The bias toward smaller formation radii is most pronounced at higher metallicities. 
Furthermore, we have argued that the migrated population also naturally forms a thick disk. This thick 
disk is in many ways reminiscent of the observed thick disk in the MW, yet its formation does not 
in any way depend on the galaxy's cosmological environment. Within the next decade, we can look 
forward to ground-breaking new surveys such as PanSTARRS, LSST and GAIA that will yield vast new 
datasets, detailing the structure of our Galaxy. Combined with detailed dynamical models, such data
will allow us to finally discriminate between the various scenarios for the formation of our
MW disk. 

\bibliography{roskar_r}

\begin{thebibliography}{}
\expandafter\ifx\csname natexlab\endcsname\relax\def\natexlab#1{#1}\fi
\expandafter\ifx\csname url\endcsname\relax
  \def\url#1{\texttt{#1}}\fi
\expandafter\ifx\csname urlprefix\endcsname\relax\def\urlprefix{URL }\fi
\providecommand{\eprint}[2][]{\url{#2}}

\bibitem[{{Bensby} et~al.(2003){Bensby}, {Feltzing}, \&
  {Lundstr{\"o}m}}]{Bensby:2003}
{Bensby}, T., {Feltzing}, S., \& {Lundstr{\"o}m}, I. 2003, \aap, 410, 527

\bibitem[{{Bensby} et~al.(2005){Bensby}, {Feltzing}, {Lundstr{\"o}m}, \&
  {Ilyin}}]{Bensby:2005}
{Bensby}, T., {Feltzing}, S., {Lundstr{\"o}m}, I., \& {Ilyin}, I. 2005, \aap,
  433, 185

\bibitem[{{Binney}(2007)}]{Binney:2007}
{Binney}, J. 2007, {Dynamics of Disks} (Island Universes - Structure and
  Evolution of Disk Galaxies), 67

\bibitem[{{Binney} \& {Tremaine}(2008)}]{Binney:2008}
{Binney}, J., \& {Tremaine}, S. 2008, {Galactic Dynamics: Second Edition}
  (Galactic Dynamics: Second Edition, by James Binney and Scott Tremaine.~ISBN
  978-0-691-13026-2 (HB).~Published by Princeton University Press, Princeton,
  NJ USA, 2008.)

\bibitem[{{Bochanski} et~al.(2010){Bochanski}, {Hawley}, {Covey}, {West},
  {Reid}, {Golimowski}, \& {Ivezi{\'c}}}]{Bochanski:2010}
{Bochanski}, J.~J., {Hawley}, S.~L., {Covey}, K.~R., {West}, A.~A., {Reid},
  I.~N., {Golimowski}, D.~A., \& {Ivezi{\'c}}, {\v Z}. 2010, \aj, 139, 2679

\bibitem[{{Boissier} \& {Prantzos}(1999)}]{Boissier:1999}
{Boissier}, S., \& {Prantzos}, N. 1999, \mnras, 307, 857

\bibitem[{{Bullock} et~al.(2001){Bullock}, {Dekel}, {Kolatt}, {Kravtsov},
  {Klypin}, {Porciani}, \& {Primack}}]{Bullock:2001}
{Bullock}, J.~S., {Dekel}, A., {Kolatt}, T.~S., {Kravtsov}, A.~V., {Klypin},
  A.~A., {Porciani}, C., \& {Primack}, J.~R. 2001, \apj, 555, 240

\bibitem[{{Carigi}(1996)}]{Carigi:1996}
{Carigi}, L. 1996, Revista Mexicana de Astronomia y Astrofisica, 32, 179

\bibitem[{{Chiappini} et~al.(2001){Chiappini}, {Matteucci}, \&
  {Romano}}]{Chiappini:2001}
{Chiappini}, C., {Matteucci}, F., \& {Romano}, D. 2001, \apj, 554, 1044

\bibitem[{{Daflon} \& {Cunha}(2004)}]{Daflon:2004}
{Daflon}, S., \& {Cunha}, K. 2004, \apj, 617, 1115

\bibitem[{{Debattista} et~al.(2006){Debattista}, {Mayer}, {Carollo}, {Moore},
  {Wadsley}, \& {Quinn}}]{Debattista:2006}
{Debattista}, V.~P., {Mayer}, L., {Carollo}, C.~M., {Moore}, B., {Wadsley}, J.,
  \& {Quinn}, T. 2006, \apj, 645, 209

\bibitem[{{Edvardsson} et~al.(1993){Edvardsson}, {Andersen}, {Gustafsson},
  {Lambert}, {Nissen}, \& {Tomkin}}]{Edvardsson:1993}
{Edvardsson}, B., {Andersen}, J., {Gustafsson}, B., {Lambert}, D.~L., {Nissen},
  P.~E., \& {Tomkin}, J. 1993, \aap, 275, 101

\bibitem[{{Fischer} \& {Valenti}(2005)}]{Fischer:2005}
{Fischer}, D.~A., \& {Valenti}, J. 2005, \apj, 622, 1102

\bibitem[{{Ghezzi} et~al.(2010){Ghezzi}, {Cunha}, {Schuler}, \&
  {Smith}}]{Ghezzi:2010}
{Ghezzi}, L., {Cunha}, K., {Schuler}, S.~C., \& {Smith}, V.~V. 2010, \apj, 725,
  721

\bibitem[{{Holmberg} et~al.(2009){Holmberg}, {Nordstr{\"o}m}, \&
  {Andersen}}]{Holmberg:2009}
{Holmberg}, J., {Nordstr{\"o}m}, B., \& {Andersen}, J. 2009, \aap, 501, 941

\bibitem[{{Juri{\'c}} et~al.(2008){Juri{\'c}}, {Ivezi{\'c}}, {Brooks},
  {Lupton}, {Schlegel}, {Finkbeiner}, {Padmanabhan}, {Bond}, {Sesar},
  {Rockosi}, {Knapp}, {Gunn}, {Sumi}, {Schneider}, {Barentine}, {Brewington},
  {Brinkmann}, {Fukugita}, {Harvanek}, {Kleinman}, {Krzesinski}, {Long},
  {Neilsen}, {Nitta}, {Snedden}, \& {York}}]{Juric:2008}
{Juri{\'c}}, M., {Ivezi{\'c}}, {\v Z}., {Brooks}, A., {Lupton}, R.~H.,
  {Schlegel}, D., {Finkbeiner}, D., {Padmanabhan}, N., {Bond}, N., {Sesar}, B.,
  {Rockosi}, C.~M., {Knapp}, G.~R., {Gunn}, J.~E., {Sumi}, T., {Schneider},
  D.~P., {Barentine}, J.~C., {Brewington}, H.~J., {Brinkmann}, J., {Fukugita},
  M., {Harvanek}, M., {Kleinman}, S.~J., {Krzesinski}, J., {Long}, D.,
  {Neilsen}, E.~H., Jr., {Nitta}, A., {Snedden}, S.~A., \& {York}, D.~G. 2008,
  \apj, 673, 864

\bibitem[{{L{\'e}pine} et~al.(2003){L{\'e}pine}, {Acharova}, \&
  {Mishurov}}]{Lepine:2003}
{L{\'e}pine}, J.~R.~D., {Acharova}, I.~A., \& {Mishurov}, Y.~N. 2003, \apj,
  589, 210

\bibitem[{{Loebman} et~al.(2010){Loebman}, {Roskar}, {Debattista}, {Ivezic},
  {Quinn}, \& {Wadsley}}]{Loebman:2010}
{Loebman}, S.~R., {Roskar}, R., {Debattista}, V.~P., {Ivezic}, Z., {Quinn},
  T.~R., \& {Wadsley}, J. 2010, ArXiv e-prints

\bibitem[{{Matteucci} \& {Francois}(1989)}]{Matteucci:1989}
{Matteucci}, F., \& {Francois}, P. 1989, \mnras, 239, 885

\bibitem[{{Minchev} \& {Famaey}(2010)}]{Minchev:2010}
{Minchev}, I., \& {Famaey}, B. 2010, \apj, 722, 112

\bibitem[{{Navarro} et~al.(1997){Navarro}, {Frenk}, \& {White}}]{Navarro:1997}
{Navarro}, J.~F., {Frenk}, C.~S., \& {White}, S.~D.~M. 1997, \apj, 490, 493

\bibitem[{{Nordstr{\"o}m} et~al.(2004){Nordstr{\"o}m}, {Mayor}, {Andersen},
  {Holmberg}, {Pont}, {J{\o}rgensen}, {Olsen}, {Udry}, \&
  {Mowlavi}}]{Nordstrom:2004}
{Nordstr{\"o}m}, B., {Mayor}, M., {Andersen}, J., {Holmberg}, J., {Pont}, F.,
  {J{\o}rgensen}, B.~R., {Olsen}, E.~H., {Udry}, S., \& {Mowlavi}, N. 2004,
  \aap, 418, 989

\bibitem[{{Pasquini} et~al.(2007){Pasquini}, {D{\"o}llinger}, {Weiss},
  {Girardi}, {Chavero}, {Hatzes}, {da Silva}, \& {Setiawan}}]{Pasquini:2007}
{Pasquini}, L., {D{\"o}llinger}, M.~P., {Weiss}, A., {Girardi}, L., {Chavero},
  C., {Hatzes}, A.~P., {da Silva}, L., \& {Setiawan}, J. 2007, \aap, 473, 979

\bibitem[{{Portegies Zwart}(2009)}]{Portegies-Zwart:2009}
{Portegies Zwart}, S.~F. 2009, \apjl, 696, L13

\bibitem[{{Ro{\v s}kar} et~al.(2008{\natexlab{a}}){Ro{\v s}kar}, {Debattista},
  {Quinn}, {Stinson}, \& {Wadsley}}]{Roskar:2008a}
{Ro{\v s}kar}, R., {Debattista}, V.~P., {Quinn}, T.~R., {Stinson}, G.~S., \&
  {Wadsley}, J. 2008{\natexlab{a}}, \apjl, 684, L79

\bibitem[{{Ro{\v s}kar} et~al.(2008{\natexlab{b}}){Ro{\v s}kar}, {Debattista},
  {Stinson}, {Quinn}, {Kaufmann}, \& {Wadsley}}]{Roskar:2008}
{Ro{\v s}kar}, R., {Debattista}, V.~P., {Stinson}, G.~S., {Quinn}, T.~R.,
  {Kaufmann}, T., \& {Wadsley}, J. 2008{\natexlab{b}}, \apjl, 675, L65

\bibitem[{{Sch{\"o}nrich} \& {Binney}(2009{\natexlab{a}})}]{Schonrich:2009}
{Sch{\"o}nrich}, R., \& {Binney}, J. 2009{\natexlab{a}}, \mnras, 396, 203

\bibitem[{{Sch{\"o}nrich} \& {Binney}(2009{\natexlab{b}})}]{Schonrich:2009a}
--- 2009{\natexlab{b}}, \mnras, 399, 1145

\bibitem[{{Sellwood} \& {Binney}(2002)}]{Sellwood:2002}
{Sellwood}, J.~A., \& {Binney}, J.~J. 2002, \mnras, 336, 785

\bibitem[{{Twarog}(1980)}]{Twarog:1980}
{Twarog}, B.~A. 1980, \apj, 242, 242

\bibitem[{{Wadsley} et~al.(2004){Wadsley}, {Stadel}, \& {Quinn}}]{Wadsley:2004}
{Wadsley}, J.~W., {Stadel}, J., \& {Quinn}, T. 2004, New Astronomy, 9, 137

\bibitem[{{West} et~al.(2008){West}, {Hawley}, {Bochanski}, {Covey}, {Reid},
  {Dhital}, {Hilton}, \& {Masuda}}]{West:2008}
{West}, A.~A., {Hawley}, S.~L., {Bochanski}, J.~J., {Covey}, K.~R., {Reid},
  I.~N., {Dhital}, S., {Hilton}, E.~J., \& {Masuda}, M. 2008, \aj, 135, 785

\bibitem[{{Wielen} et~al.(1996){Wielen}, {Fuchs}, \& {Dettbarn}}]{Wielen:1996}
{Wielen}, R., {Fuchs}, B., \& {Dettbarn}, C. 1996, \aap, 314, 438

\bibitem[{{Wyse}(2008)}]{Wyse:2008}
{Wyse}, R.~F.~G. 2008, ArXiv e-prints

\end{thebibliography}

\end{document}